# HIGHER SCHOOL OF ECONOMICS
NATIONAL RESEARCH UNIVERSITY

*Fuad Aleskerov, Elizaveta Victorova*

## AN ANALYSIS OF POTENTIAL CONFLICT ZONES IN THE ARCTIC REGION

As a result of the climate change the situation in Arctic area leads to several important consequences. On the one hand, oil and gas resources can be exploited much easier than before. Thus, one can already observe discussions on disputed shelf zones where the deposits are located. On the other hand, oil and gas excavation leads to serious potential threats to fishing by changing natural habitats which in turn can create serious damage to the economies of some countries in the region. Another set of problems arises due to the extension of navigable season for Arctic Shipping Routes.

We present a model allowing one to analyze preferences of the countries interested in natural, maritime and fish resources, and reveal potential conflicts among them. We strongly believe that early forecast of such potential conflict zones might ease the decision making process in international relations.

*Aleskerov Fuad* – National Research University Higher School of Economics and Institute of Control Sciences of Russian Academy of Sciences.

*Victorova Elizaveta* – National Research University Higher School of Economics.

# Introduction

The area north of the Arctic Circle is known as the Arctic Region. Sea and glacier ice in the Arctic is rapidly melting making the Arctic waters more accessible for resource exploration and exploitation. Global demand for oil and gas is also increasing although there are attempts to switch to other energy sources. Increasing interest in the Arctic Region is driven by the volume of natural resources concentrated in the area and rapid development of new technologies allowing their exploitation. Apart from that due to ice melting fishing seasons are considerably extended by increased periods of open water, as are the opportunities for Arctic Shipping Routes which reduce transit distances between Europe and Asia, Asia and Northern America [13, 15].

The five Arctic States – Canada, Denmark, Norway, Russia and the USA – are limited to an exclusive economic zone (EEZ) of 200 nautical miles adjacent to their coasts. The waters beyond the EEZs are considered the "high seas" or international waters. The international waters are not owned by any country [19].

Today many counties apart from Arctic States are showing their interest to the region. The Arctic Council reflects international interests. It consists of Canada, Denmark (including Greenland and the Faroe Islands), Finland, Iceland, Norway, Russia, Sweden and the United States. Moreover, twelve non-Arctic countries have been admitted as observers to the Arctic Council (France, Germany, the Netherlands, Poland, Spain, United Kingdom, China, Italy, Japan, Republic of Korea, Singapore, India) [4].

"Upon ratification of the United Nations Convention on the Law of the Sea (UNCLOS), a country has a ten-year period to make claims to an extended continental shelf which, if validated, gives it exclusive rights to resources on or below the seabed of that extended shelf area" [19]. Canada, Denmark, Norway, Russian Federation and the USA have already made several claims. The decision can have dramatic impact on countries economy. If the Commission on the Limits of the Continental Shelf recognizes a zone as a part of the countries continental shelf, then no other country can have access to its mineral and fish resources.

The situation raises the question of identifying areas of potential conflict of interests. This paper looks at the countries interests based on their agenda.

The structure of the paper: Section 1 gives some information about the countries that have shown explicit interest in the Arctic Region. Section 2 describes the model analyzing mutual interests. Section 3 presents the results. Section 4 concludes.

## 1. Geographic information about the countries

Arctic States include eight countries five of which have a direct access to the Arctic Ocean. In our research we focus on Canada, Denmark (Greenland), Iceland, Norway, Russia and the United States as their EEZs capture Arctic waters. Non-Arctic countries under consideration are China, Japan and South Korea. Figure 1 illustrates the countries and Table 1 provides general information.

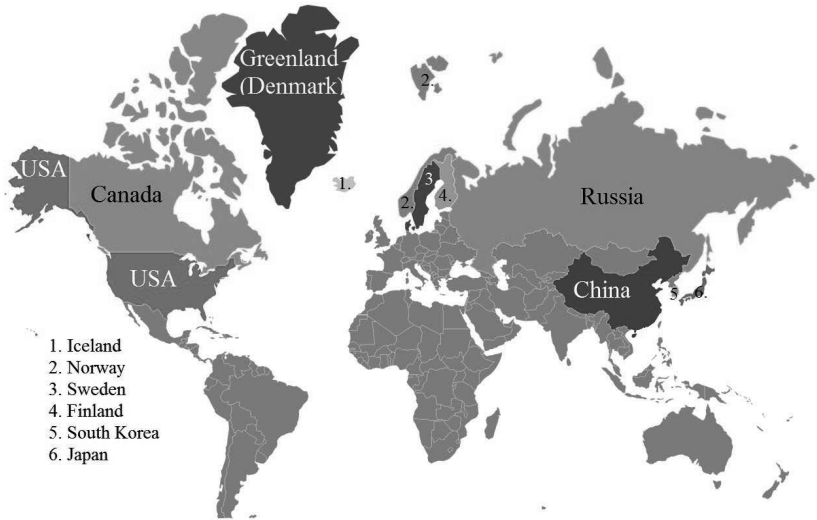

**Figure 1.** Arctic and non-Artic States

*Table 1.* General information about the countries

| Country | Region | Area, thousand km² | Population, m | GDP, $bn | GDP, $ thousand per capita |
|---|---|---|---|---|---|
| Canada | Artic | 9 900 | 36 | 1 518 | 43 |
| Denmark | Arctic | 2 210 | 5.7 | 347 | 44 |
| Finland | Arctic | 338 | 5.4 | 276 | 50 |
| Iceland | Arctic | 103 | 0.3 | 14 | 44 |
| Norway | Arctic | 385 | 5.1 | 353 | 68 |
| Russia | Arctic | 17 000 | 144 | 3 559 | 24 |
| Sweden | Arctic | 450 | 9.7 | 573 | 58 |
| USA | Arctic | 9 800 | 320 | 16 720 | 52 |
| China | Non-Arctic | 9 600 | 1 350 | 16 150 | 12 |
| Japan | Non-Arctic | 377 | 126 | 4 800 | 38 |
| Korea | Non-Arctic | 100 | 51 | 1 790 | 35 |

**1.1. Arctic States**

According to International Energy Agency, Russia, USA and Canada are in the top five oil-producing countries. USA, Russia, Canada and Norway are also in the top six gas-producing countries in the world. Moreover, Russia, Norway and Canada are leading petrol and gas exporters. It is worth mentioning that five out of six Arctic States under consideration (except Iceland) have offshore mines. Exploiting more Arctic deposits can become a new source of income for these countries.

Another source of profit in the Region is the maritime activity. To navigate through the Northern Sea Route (NSR), which lies within Russian EEZ, one must obtain official permission and pay a fee that includes inspection of the ship for ice navigation worthiness as well as the necessary icebreaker and pilotage support. In contrast, the Northwest Passage (NWP) does not have a formal transit system. It is due to the fact that navigation through NWP is extremely challenging. However, in light of ice-melting Canadian Government is interested in economic development of the area in the future [9, 11].

Economic structure of Iceland and Greenland (Denmark) differs from that of the aforementioned countries. These two States are highly dependent on the fishing. Thus, preserving natural habitats and controlling the amount of capture is vital. In addition, increase in resource developments and cargo trafficking worsens the situation in terms of pollution making the States interested in controlling energy and maritime resources.

**1.2. Non-Arctic States**

Non-Arctic countries under consideration are China, Japan and South Korea. All three countries show growing interest in the Arctic Region. It concerns not only resources but also scientific research. Thus, Japan has a fifty-year-old history in polar research, Japan and China established Institutes specialized in such research in early 90s and early 2000s, respectively. Republic of Korea is also an active researcher and has had an operating Dasan Arctic research station on the Svalbard, Norway, since 2002 [12].

Although the countries base their involvement on UNCLOS it is in their best interest that international waters remain international. In contrast to Arctic States, peripheral actors can demand permission to exploit resources only if the area of exploitation does not belong to any country, thus, making it common heritage of mankind [19].

Over the last two decades China's, Japan's and Korean gas and oil import has increased. Large share of those resources comes from Middle East. In more detail, half of China's oil, about 85 percent of oil and one third of Japan's gas and 62% of oil and 53% of Korean gas are imported through the Malacca Strait. There are certain concerns about the energy import's vulnerability in a case of the Straits sudden shutdown. This makes the three countries interested in the Arctic deposits and routes.

One more concern is the piracy. Rising threats has made ship-insurance exorbitantly costly. Some companies made a decision to sail around the southern tip of Africa to avoid risks of navigating towards the Suez Canal. Sailing through the NWP can become an economically attractive substitution [16].

## 2. Model description

We model countries preferences via utility functions. A utility function is a matrix with elements representing certain area as shown on Figure 2. For each area the utility of each resource is evaluated based on countries interest.

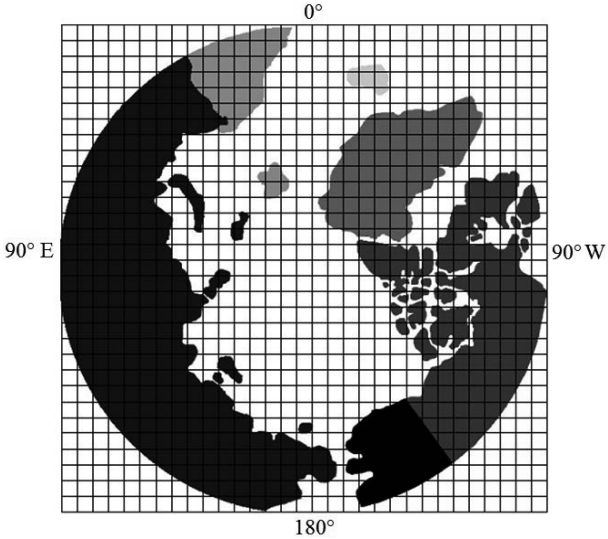

**Figure 2.** Areas for which utility functions are constructed

### 2.1. Information Sources

Although the interests in the given resource may be contradictory (e.g., one can want to increase or decrease its extraction in the area for various reasons) the most valuable ones remain the same: oil, gas, fish and maritime resources [10, 20].

In 2008 a team of U.S. Geological Survey (USGS) gauged the possibility of discovering oil and gas reserves in the future. The appraisal evaluated the petroleum potential of the provinces of Arctic Region (to the north of 66° north latitude). Figures 3 and 4 show the results of the appraisal that are used in our research [5].

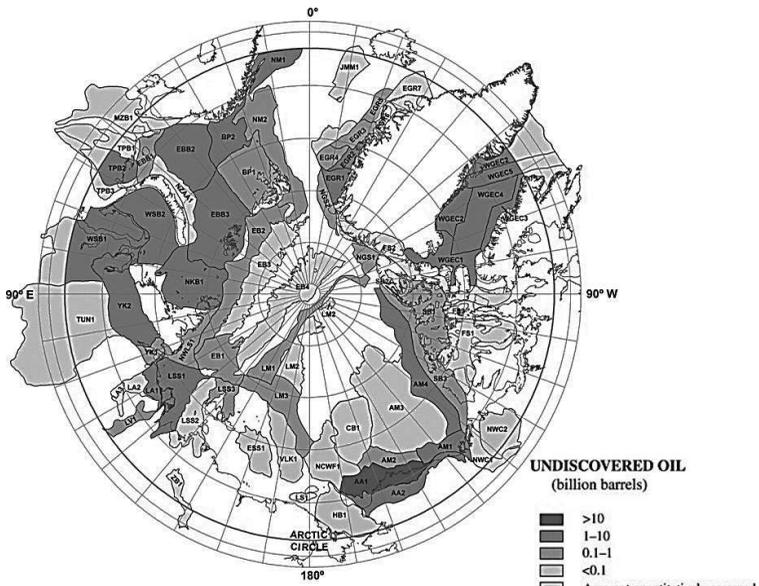

**Figure 3.** Undiscovered oil [5]

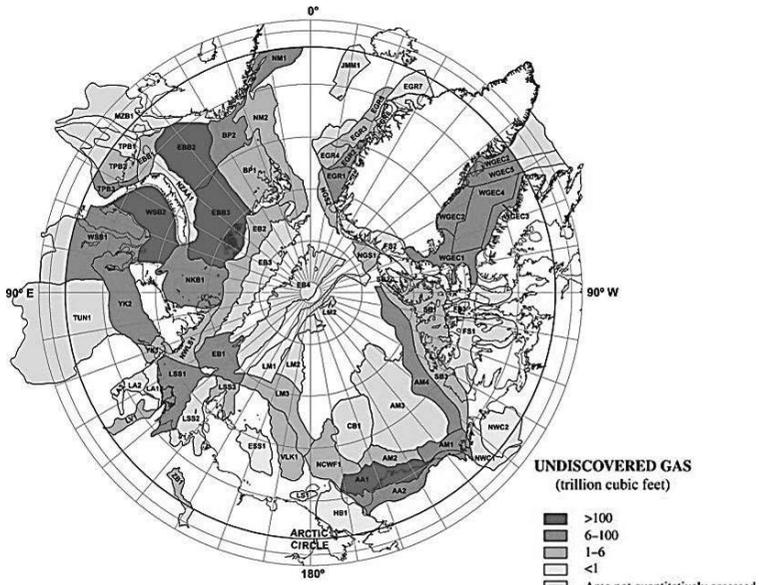

**Figure 4.** Undiscovered gas [5]

Construction of maritime utility function was based on a simulation of optimal navigation routes for current conditions (2006–2015) and expected conditions (2040–2059). Figure 5 illustrates these routes for years 2040–2059 [17].

**Figure 5.** Optimal navigation routes in 2040–2059

Finally, the fish utility function was based on the information about the year-round range of some fish species that is caught commercially.

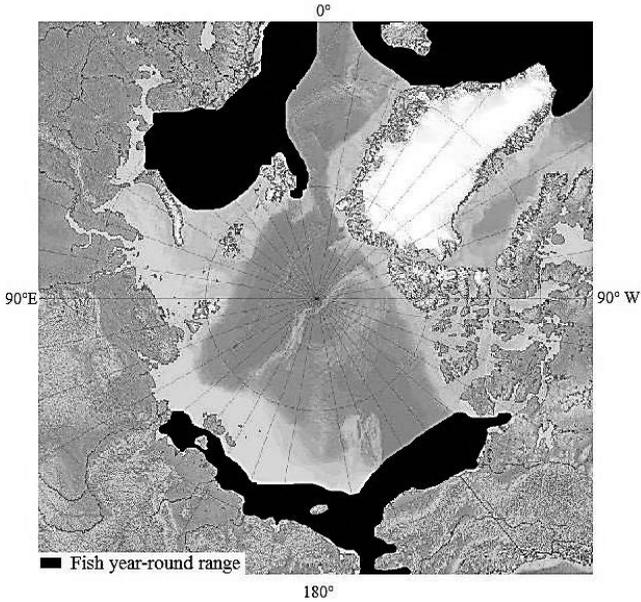

**Figure 6.** Fish year-round range

## 2.2 Construction of the Utility Functions

Based on the input data the utility functions are estimated for every country under consideration. First of all an intermediary distance function is constructed for every country. This function $dist_{country}(area)$ is equal to the distance between the *country* and the *area*:

$$dist_{country}(area) = \begin{cases} 0, & area \in country \\ \text{distance in km}, & area \notin any\ country \\ \infty, & area \in other\ country \end{cases}.$$

As the resources are different the following assumptions are made:
1. Gas, petroleum and fish reserves are of interest for all countries:

$$f_{country,gas}(area) = \frac{Gas\ deposit(area)}{g\left(dist_{country}(area)\right)},$$

where $g(*)$ is a function and *Gas deposit(area)* is the size of the deposit. Function $g(*)$ can differ from country to country.

The same formula applies for oil and fish utility function.

2. Maritime resources differ from the others because the distance does not matter that much in this case. The utility of a route for a transiting company as well as non-Arctic State stays the same lengthwise. However, Arctic States are mostly interested in the routes that lie within their EEZs.

$$f_{Arctic\ Country,maritime}(area) = \begin{cases} a, & area \in country's\ EEZ \\ \dfrac{a}{h(distance\ in\ km)}, & area \notin any\ country \end{cases},$$

$$f_{non-Arctic\ Country,maritime}(area) = a \cdot \text{Importance}_{non-Arctic\ Country}$$

where *area* belongs to one of the shipping routes, $h(*)$ is a function and the $\text{Importance}_{non-Arctic\ Country}$ is a constant that depends on how much the country could save if sailing along Arctic Shipping Routes, $\text{Importance}_{non-Arctic\ Country} \leq 1$.

Naturally, other forms of utility functions can be applied.

## 2.3. An aggregation procedure

After separate utility functions are constructed the problem of their aggregation allowing comparison of mutual interests in different activities arises. Aggregated function reflects intensity of a potential conflict in the areas. The functions are made discrete on n-grade scale. We use a 6-grade scale, where 0 reflect no interest in the area and 5 – extremely high interest.

Let us consider gas utility functions. They are described by 9 matrices that indicate how important each element is for each country. Each element $(i,j)$ is represented by a vector $f_{gas} = f_{gas}(i,j) = \left( f_{C_1}(i,j), \ldots, f_{C_9}(i,j) \right)$.

To compare the intensity of a potential conflict of interests we use threshold aggregation rule described in [1–3]. Let $I(i,j)$ be the intensity of a conflict of interests in the area $(i,j)$. To compare $I_1 = I(i_1, j_1)$ and $I_2 = I(i_2, j_2)$ we use lexicographic ordering. To apply it vector components are sorted so that $f_{gas} = \left( f_{C_1}, \ldots, f_{C_9} \right) = \left( f_{C_{s1}}, \ldots, f_{C_{s9}} \right), f_{C_{s1}} \geq f_{C_{s2}} \ldots \geq f_{C_{s9}}$, where $C_{s1}, \ldots, C_{s9}$ is a permutation of indices $1, \ldots, 9$. Then inverse lexicographic ordering is applied, i.e. the first 'letter' is the biggest number and the last is the smallest. We use 6-grade scale, so the first 'letter' is 5 and the last is 0.

To divide elements into sets that represent different intensities of conflicts thresholds $T_1, \ldots T_N$ are used. If the intensity $I$ in the area $(i,j)$ is $T_k \succ I \succ T_{k+1}$, then the area belongs to a set $k$.

Figure 7 presents the whole aggregation procedure. First, the described rule is applied to each of the resources separately. As a result four forecasts are constructed based on a single resource. Then the rule is applied to the forecasts to assess the intensity of the overall conflicts of interest.

It is important to note that different parameters can be introduced to the aggregation rule. For example, the extent $\alpha$ to which a country is interested in the EEZs of other countries can alter utility functions. If $\alpha = 0$, then the country's interests lie within the international waters and it's EEZ only. However, that does not reflect the reality. Country's ecology and economy depend on its neighbors. Setting the safety and drilling standards and fish quotas are international matters and thus $\alpha > 0$ for all countries. If $\alpha = 1$, the country considers all EEZs international. When developing a given scenario, parameter $\alpha_{country, resource}$ should be introduced for every country and for each resource. Other important parameter is a weight of the resource. If some country considers one resource more important than others, then the resource has a higher weight. This can be considered on both steps: for every single country (step 1) and as a global tendency (step 2).

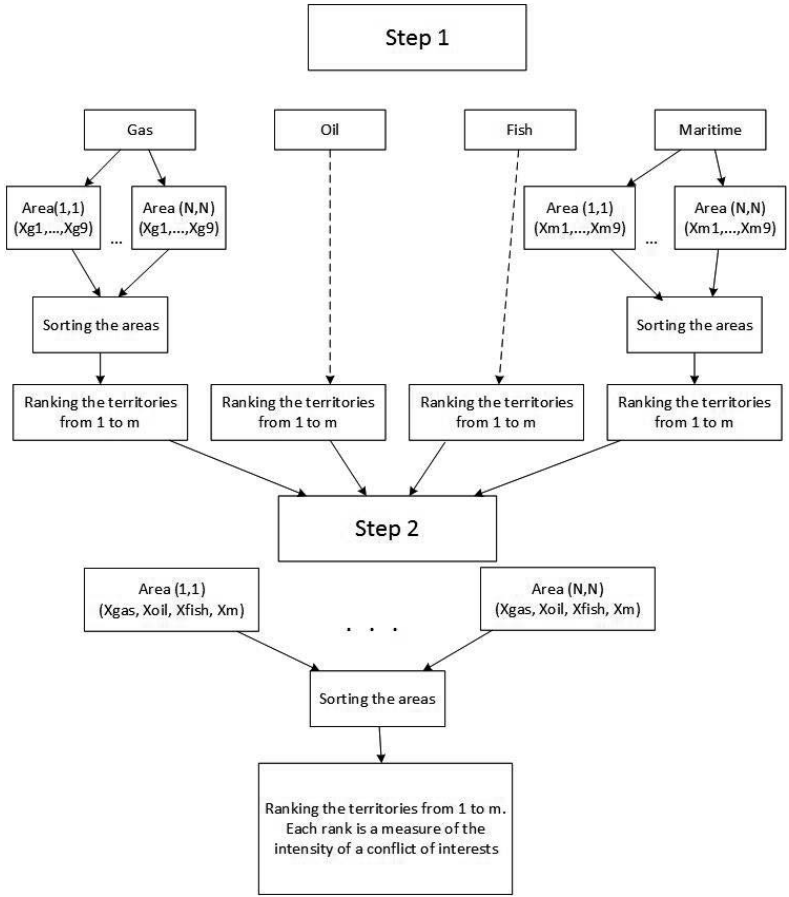

**Figure 7.** Aggregation procedure

## 3. Results

First, let us consider that the only difference between different scenarios is a value of a parameter $\alpha$. Figure 8 shows how the change in $\alpha$ effects the resulting intensities of potential conflict of interests. As $\alpha$ increases, the intensities shift from the international area to EEZs of the Arctic States. Nevertheless, all the results highlight Barents Basin as a tense subregion.

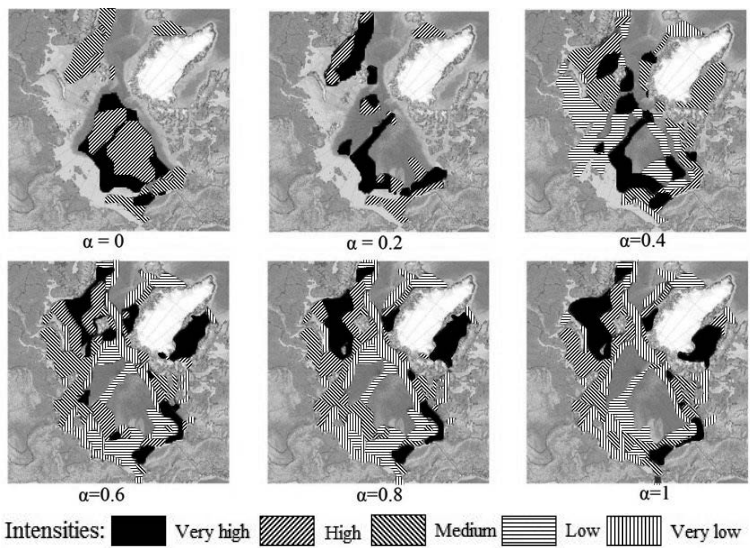

**Figure 8.** Results depending on $\alpha$

Now let us consider cases of high and low interest. In the former case non-Arctic country's interests are much higher than in the latter, $\alpha$ is equal to 0.8 and 0.1 respectively.

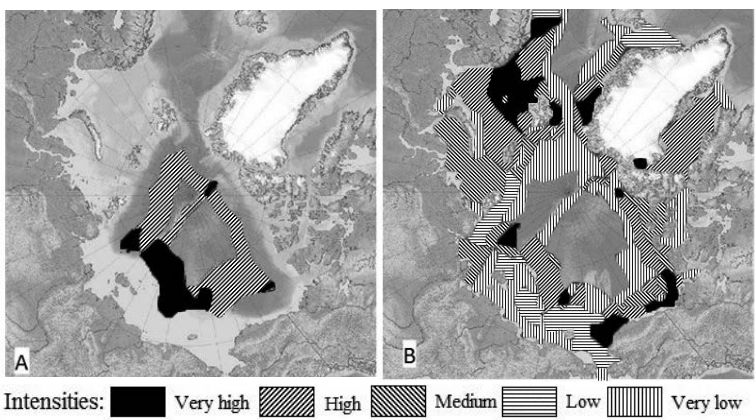

**Figure 9.** Scenarios of A) low and B) high mutual interests.

In a low interest case most area's intensities are very low. Territories with two higher ranks are concentrated around the Pole. High interests lead to increase in the intensities of a conflict of interests.

## 4. Conclusion

We have assessed the intensity of potential conflict of interests in the Arctic Region based on some parameters. Those parameters, however, are expert judgments which should be studied separately.

The constructed model is a prototype. It needs further development in several ways. First, publically available data should be replaced by more accurate results of geological (gas and oil), biological (fish) and economic (maritime resources) surveys. Then, the evaluation of parameters should be developed. Finally, introducing more detailed scenarios can be implemented.

The next step should be the development of game-theoretic model predicting and analyzing equilibria ruling the behavior of the countries and their optimal policies in the Arctic area.

**Acknowledgement**
The article was prepared within the framework of the Basic Research Program at the National Research University Higher School of Economics (HSE). We are grateful to the International Laboratory of Decision and Choice Analysis of the National Research University Higher School of Economics for partial financial support. We thank Professor Panos Pardalos for the interest to and support of this work.

В результате изменения климата ситуация в Арктике серьезно меняется. С одной стороны, месторождения нефти и газа могут разрабатываться гораздо проще, чем раньше. Именно поэтому уже можно слышать о спорных зонах на шельфе, где располагаются запасы этих ресурсов. С другой стороны, добыча нефти и газа ведет к серьезным угрозам для рыбных ресурсов, что может иметь разрушительные последствия для экономик ряда стран в регионе. Другой пласт проблем связан с удлинением навигационного периода в арктических судоходных артериях.

Мы предлагаем модель, позволяющую анализировать предпочтения стран в отношении полезных ископаемых, рыбных ресурсов и судоходных путей, и выявлять зоны потенциальных конфликтов между странами.

Мы убеждены, что раннее выявление таких зон может сильно упростить процесс принятия решений в международных отношениях.

Ключевые слова: Арктический регион, залежи нефти и газа, рыбные ресурсы, арктические судоходные пути, зоны конфликтов, многокритериальная оценка, пороговое агрегирование

*Алескеров Ф.Т.* – Национальный исследовательский университет «Высшая школа экономики»; Институт проблем управления РАН.

*Викторова Е.Р.* – Национальный исследовательский университет «Высшая школа экономики».

Алескеров Фуад Тагиевич, Викторова Елизавета Романовна

# Анализ зон потенциальных конфликтов в Арктическом регионе

(*на английском языке*)